\documentclass[reprint, prl, aps, superscriptaddress, twocolumn, showkeys]{revtex4-1}
\usepackage{amssymb}
\usepackage{amsmath}
\usepackage{graphicx}
\usepackage{bm}
\usepackage{dcolumn}
\usepackage{color}
\usepackage{ulem}
\usepackage{float}
\usepackage{textcomp}
\restylefloat{table}
\usepackage{verbatim}
\usepackage{epstopdf}
\usepackage[colorlinks]{hyperref}
\usepackage[mediumspace,mediumqspace,squaren]{SIunits}

\begin{document}

\title{Valley-tunable, even-denominator fractional quantum Hall state in the lowest Landau level of an anisotropic system}
\date{\today}

\author{Md.\ Shafayat Hossain}
\author{Meng K.\ Ma}
\author{Y. J.\ Chung}
\author{S. K.\ Singh} 
\author{A. \ Gupta} 
\author{K. W.\ West}
\author{K. W.\ Baldwin}
\author{L. N. \ Pfeiffer}
\affiliation{Department of Electrical and Computer Engineering, Princeton University, Princeton, New Jersey 08544, USA}
\author{R.\ Winkler}
\affiliation{Department of Physics, Northern Illinois University, DeKalb, Illinois 60115, USA}

\author{M.\ Shayegan}
\affiliation{Department of Electrical and Computer Engineering, Princeton University, Princeton, New Jersey 08544, USA}

%%%%%%%%%%%%%%%%% END OF PREAMBLE %%%%%%%%%%%%%%%%

\begin{abstract}

Fractional quantum Hall states (FQHSs) at even-denominator Landau level filling factors ($\nu$) are of prime interest as they are predicted to host exotic, topological states of matter. We report here the observation of a FQHS at $\nu=1/2$ in a two-dimensional electron system of exceptionally high quality, confined to a wide AlAs quantum well, where the electrons can occupy multiple conduction-band valleys with an anisotropic effective mass. The anisotropy and multi-valley degree of freedom offer an unprecedented tunability of the $\nu=1/2$ FQHS as we can control both the valley occupancy via the application of in-plane strain, and the ratio between the strengths of the short- and long-range Coulomb interaction by tilting the sample in the magnetic field to change the electron charge distribution. Thanks to this tunability, we observe phase transitions from a compressible Fermi liquid to an incompressible FQHS and then to an insulating phase as a function of tilt angle. We find that this evolution and the energy gap of the $\nu=1/2$ FQHS depend strongly on valley occupancy. 

\end{abstract}

\maketitle

%Fractional quantum Hall states (FQHSs) at even-denominator Landau level filling factors ($\nu$) are widely believed to host a novel class of matter that obeys non-Abelian statistics and could be of potential use in topological quantum computing. We report here the observation of a FQHS at $\nu=1/2$ in a two-dimensional electron system of exceptionally high quality, confined to a wide AlAs quantum well, where the electrons can occupy multiple conduction-band valleys with an anisotropic effective mass. The anisotropy and multi-valley degree of freedom offer an unprecedented tunability of the $\nu=1/2$ FQHS as we can control both the valley occupancy via the application of in-plane strain, and the ratio between the strengths of the short- and long-range Coulomb interaction by tilting the sample in the magnetic field to change the electron charge distribution. Thanks to this tunability, we observe phase transitions from a compressible Fermi liquid to an incompressible FQHS and then to an insulating phase as a function of tilt angle. We find that this evolution and the energy gap of the $\nu=1/2$ FQHS depend strongly on valley occupancy. 

The energy dispersion of a two-dimensional electron system (2DES) is quenched into flat bands, namely Landau levels (LLs), when subjected to a perpendicular magnetic field ($B_{\perp}$). The resulting dominance of the Coulomb interaction leads to different classes of correlated electron states depending on the LL index. In the lowest ($N=0$) LL, the FQHSs are generally observed at odd-denominator LL filling factors ($\nu$) on the flanks of $\nu=1/2$ \cite{Jain.2007, Halperin.Jain.2020}. These FQHSs can be effectively described as the integer QHSs of composite fermions (CFs), weakly-interacting quasi-particles emergent from pairing each electron with two flux quanta. At $\nu=1/2$, the electron-flux attachment leads to a zero effective magnetic field for the CFs and no FQHS is observed; instead the CFs form a Fermi sea \cite{Halperin.PRB.1993, Halperin.Jain.2020, Jain.2007}. In the $N=1$ LL, on the other hand, the node in the wavefunction softens the short-range component of the Coulomb repulsion, allowing the CFs to pair up and form a Bose-Einstein condensate-type ground state \cite{Willett.PRL.1987, Willett.RPP.2013, Falson.Nat.Phys.2015, banerjee.2018, ShafayatAlAs.PRL.2018, willett.2023}. A prime example is the $\nu=5/2$ FQHS, observed in high-quality GaAs 2DESs \cite{Willett.PRL.1987, Willett.RPP.2013}, which is theoretically predicted to be  a spin-polarized (one-component), Pfaffian state \cite{Moore.Nucl.Phys.B.1991} with non-Abelian quasiparticles, and be of potential use in topological quantum computing \cite{Nayak.Rev.Mod.Phys.2008}. The full spin polarization of the $5/2$ FQHS has been confirmed in several experiments \cite{ Pan.SSC.2001, Wurstbauer.PRL.2012, Tiemann.Science.2012, Stern.PRL.2012, Eisenstein.PRL.2017, Shafayat5/2.PRL.2018}, and there is also experimental evidence suggesting that it is a Pfaffian state \cite{Willett.RPP.2013, banerjee.2018}.

\begin{figure}[t!]
\includegraphics[width=.45\textwidth]{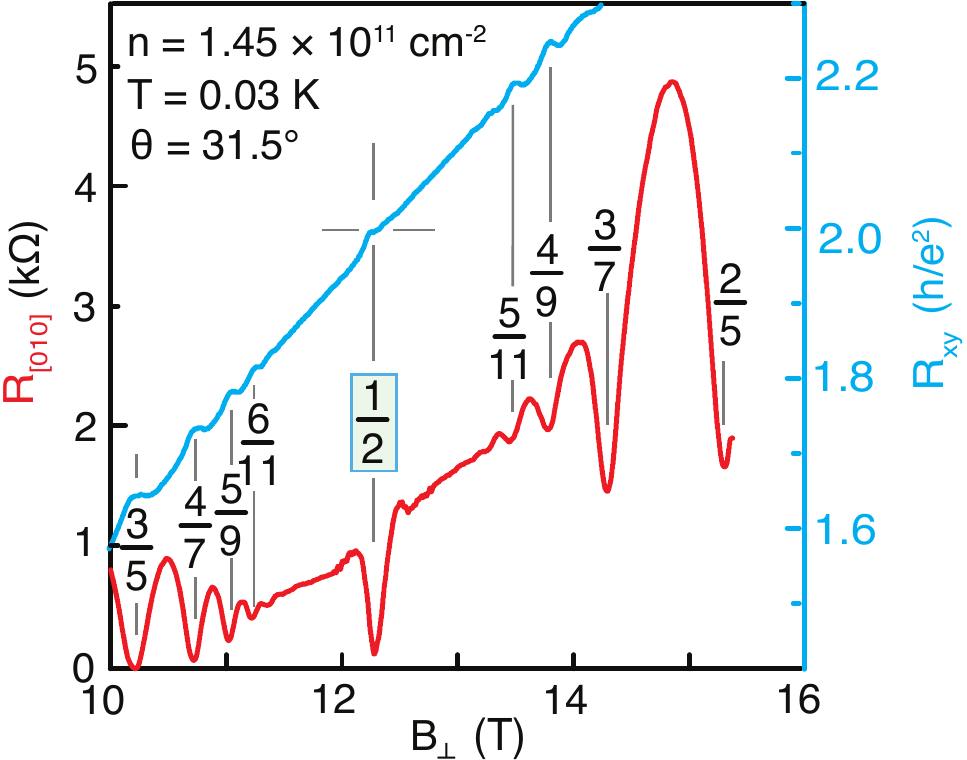}
\caption{\label{fig:Fig2} 
Observation of FQHS at $\nu=1/2$ in an X-valley occupied AlAs 2DES when subjected to a tilted magnetic field; $\theta$ denotes the angle between $B_\perp$ and total $B$. }
\end{figure} 

%Inset: calculated charge distribution at $B_{\perp}=0$ and $B_{||}$ at $\nu=1/2$ for $\theta = 31^o$.

\begin{figure*}[t!]
\includegraphics[width=.99\textwidth]{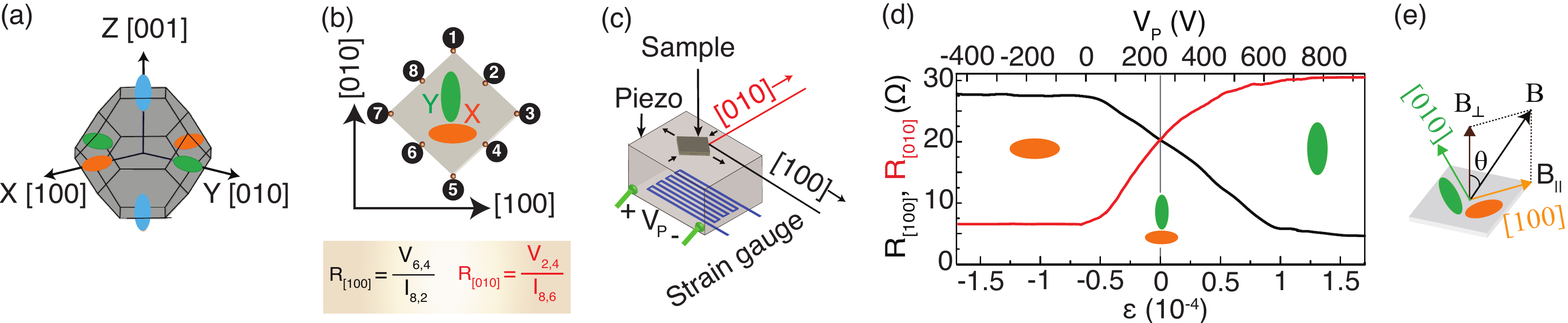}
\caption{\label{fig:Fig2} 
Sample description and valley tuning. (a) First Brillouin zone of bulk AlAs, showing anisotropic conduction band valleys. (b) Sample geometry, showing the orientation of the two occupied valleys (X and Y) and the measured resistances ($R_{[100]}$ and $R_{[010]}$).  Electrical contacts to the sample are denoted by 1-8. For $R_{[100]}$, we pass current from contact 8 to 2 and measure the voltage between contacts 6 and 4. For $R_{[010]}$, the current is passed from 8 to 6 and we measure the voltage between 2 and 4.  (c) Experimental setup for applying in-plane strain ($\varepsilon$). (d) Resistance of the sample at $B = 0$ and $T \simeq 0.03$ K, measured as a function of $\varepsilon$. (e) Direction of tilted magnetic field with respect to the crystallographic directions.} 
\end{figure*}

%(d) Magneto-transport traces for different valley occupancy showing numerous odd-denominator FQHSs. Plotted are the $R_{[100]}$ (black), $R_{[010]}$ (Red) and $R_{xy}$ (blue) traces, taken at $T\simeq 0.03$ K and at three strain values as indicated. Insets show the corresponding LL diagrams determined from the g-factor measurement as described in the Supplemental Material \cite{Suppl.Mat.} .

In the $N=0$ LL, ordinarily the Coulomb interaction dominates and leads to a compressible state at $\nu=1/2$ \cite{Halperin.Jain.2020, Halperin.PRB.1993, Jain.2007}. However, when the electron layer thickness is increased by widening the quantum well (QW), the short-range Coulomb repulsion relaxes. This opens up the possibility for CF pairing and thus an even-denominator FQHS. A FQHS at $\nu=1/2$ has indeed been reported in 2DESs confined to wide GaAs QWs where the charge distribution is bilayer-like but there is substantial interlayer tunneling \cite{Suen.PRL.1992, Suen.PRL.1994, Suen2.PRL.1992, Shabani.PRB.2013, Liu.PRL.2014, Mueed.PRL.2015b, Mueed.PRL.2016}. Its origin, however, is still under debate. Some of its aspects are consistent with a two-component, $\psi_{331}$, Halperin-Laughlin (Abelian) state \cite{Halperin.Helv.Phys.Acta.1983, Chakraborty.PRL.1987, Yoshioka.PRB.1989, Macdonald.SSc.1990, He.PRB.1993, Peterson.PRB.2010, Thiebaut.PRB.2015,  Suen.PRL.1994, Shabani.PRB.2013}. Recent experiments \cite{Mueed.PRL.2015b, Mueed.PRL.2016} and theories \cite{Zhu.PRB.2016, Faugno.PRL.2019, Zhao.PRB.2021}, on the other hand, argue strongly in favor of a one-component, Pfaffian state, in agreement with an early theoretical description \cite{Greiter.PRB.1992, Footnote.LL.crossing}. 

Here we report the observation of a $\nu=1/2$ FQHS in a wide AlAs QW where there are multiple, anisotropic conduction-band valleys whose occupancy can be tuned continuously via the application of \textit{in situ} strain.As highlighted in Fig. 1, we observe this state when the sample is tilted in the magnetic field. We demonstrate its evolution as we control the valley which the 2DES occupies, and the shape of the charge distribution as we tilt the sample.

\begin{figure*}[t!]
\includegraphics[width=.99\textwidth]{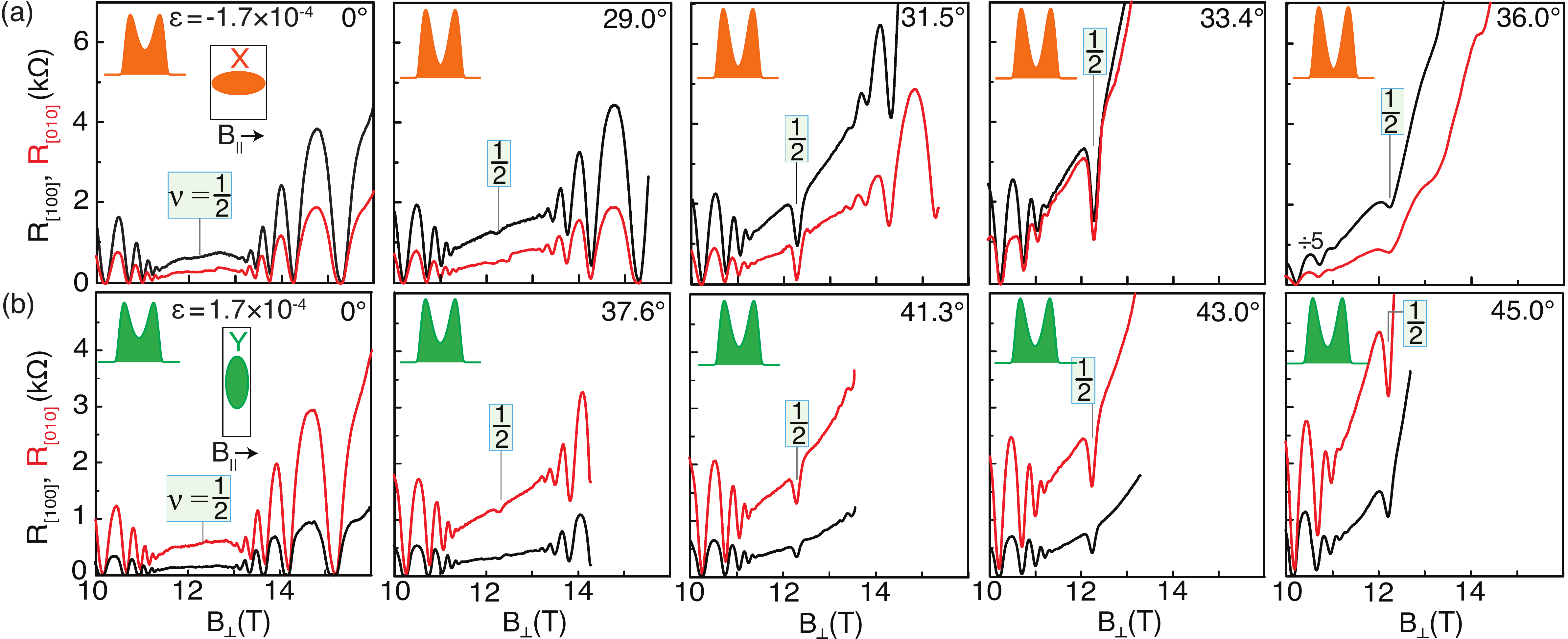}
\caption{\label{fig:Fig2} 
Tilt-evolution of magnetoresistance data near $\nu=1/2$ for different valley occupancies, all traces taken at $T\simeq 0.03$ K, and the tilt angles are indicated in each panel. The direction of $B_{||}$ is along [100], i.e., along $R_{[100]}$. (a) and (b) contain data for cases when the electrons occupy only X or only Y valley, respectively. In both (a) and (b), with increasing $\theta$, the 2DES at $\nu=1/2$ undergoes transitions from a compressible phase to an incompressible FQHS, and then finally to an insulating phase.  However, the transitions happen at relatively smaller $\theta$ for case (a) where the electrons occupy the X valley. Insets in each panel show the calculated charge distribution for the corresponding $B_{||}$ that is experienced by the electrons at $\nu=1/2$ \cite{F1}. 
}
\end{figure*} 

\begin{figure*}[t!]
\includegraphics[width=.99\textwidth]{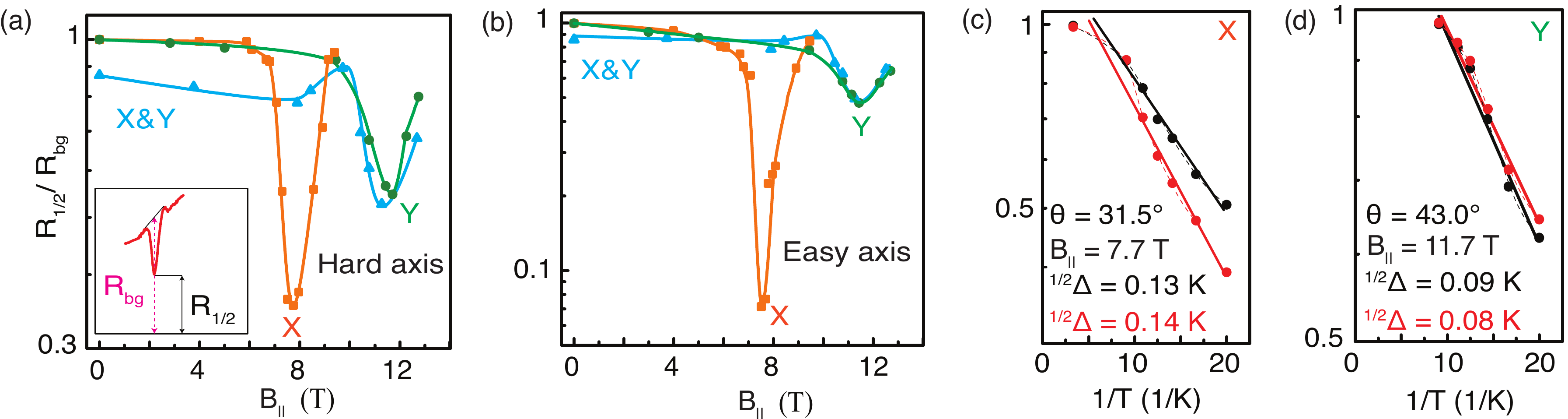}
\caption{\label{fig:Fig3} 
Strength of the $\nu=1/2$ FQHS, as defined in the inset to (a), plotted against $B_{||}$ for different valley occupancies. (a, b) Data for the hard and easy axis directions (see text). (c, d) Extraction of the pseudo activation gap ($^{1/2}\Delta$) of the $\nu=1/2$ FQHS from $T$-dependence of $R_{[100]}$ (black) and $R_{[010]}$ (red) for the two valleys. }
\end{figure*}

%\begin{figure*}[t!]
%\includegraphics[width=.97\textwidth]{fig4.pdf}
%\caption{\label{fig:Fig4} 
%Temperature dependence of the magnetoresistance for $N=0$ LL where the 2D electrons occupy only (a) X valley (b) Y valley. Plotted are the $R_{[100]}$ (black), $R_{[010]}$ (red), and $R_{xy}$ (blue) for six different $T$ ranging from $\simeq$ 0.03 to 0.3 K (c) Extraction of pseudo activation gap of $\nu=1/2$ FQHS from $T$-dependence of $R_{[100]}$ and $R_{[010]}$ for both valleys.}
%\end{figure*}

%(e) Calculated electron Fermi sea distortions, resulting from $B_{||}$ applied along [100],  for two different valleys occupancies. For clarity, in all panels, the insets show only one of the split Fermi seas. 

Our material platform is a 2DES, with density $n = 1.45 \times 10^{11}$ cm$^{-2}$ and mobility $\mu = 7.5 \times 10^{5}$ cm$^{2}/$Vs, confined to a 45-nm-wide AlAs QW. Electrons in bulk AlAs occupy three ellipsoidal valleys (X, Y, and Z), centered at the X points of the Brillouin zone, with their major axes lying in the [100], [010], and [001] crystallographic directions [Fig. 2(a)]. However, this three-fold degeneracy is lifted when we grow an AlAs QW on a GaAs substrate because the biaxial, in-plane compression in the AlAs layers originating from the lattice mismatch between AlAs and GaAs pushes the Z valley higher in energy relative to X and Y \cite{Shayegan.AlAs.Review.2006, Lay, Shayegan.APL.2003, Depoortere.APL.2002, Chung.PRM.2018}; we denote the growth direction as [001]. The 2D electrons therefore occupy only valleys X and Y, with their major axes lying in the plane {along [100] and [010], respectively [Fig. 2(b)] \cite{Shayegan.AlAs.Review.2006, Lay, Shayegan.APL.2003, Depoortere.APL.2002, Chung.PRM.2018}. Each valley possesses an anisotropic Fermi sea with longitudinal and transverse electron effective masses of $m_l = 1.1m_0$ and $m_t = 0.20m_0$, where $m_0$ is the free electron mass. 

We can break the degeneracy between the X and Y valleys and control their relative occupancy by applying an in-plane, uniaxial strain $ \varepsilon=\varepsilon_{[100]}-\varepsilon_{[010]}$, where $\varepsilon_{[100]}$ and $\varepsilon_{[010]}$ are the strain values along [100] and [010] \cite{Shayegan.APL.2003, Shayegan.AlAs.Review.2006}. This is achieved by gluing the sample to a piezo-actuator [Fig. 2(c)], and applying a voltage bias ($V_P$) to its leads to control the amount of strain \cite{Shayegan.APL.2003, Shayegan.AlAs.Review.2006}.  Figure 2(d) demonstrates how we tune and monitor the valley occupancy. Here we show the sample's piezoresistance as a function of $\varepsilon$ [measurement configurations are shown in Fig. 2(b)]. When $\varepsilon=0$, the two valleys are degenerate and equally occupied \cite{FN}, and the 2DES exhibits isotropic transport, namely, the resistances measured along [100] and [010] ($R_{[100]}$ and $R_{[010]}$) are equal, even though the individual valleys are anisotropic.  For $\varepsilon>0$, as electrons transfer from X to Y, $R_{[100]}$ decreases [black trace in Fig. 2(d)] because the electrons in Y have a small effective mass and therefore higher mobility along [100]. (Note that the total 2DES density remains fixed as strain is applied; see Fig. S8 of the Supplemental Material \cite{Suppl.Mat.}.) $R_{[100]}$ eventually saturates at a low value, when all electrons are in Y \cite{Shayegan.AlAs.Review.2006, Gokmen.Natphy.2010, ShafayatAlAs.PRL.2018}. For $\varepsilon<0$, $R_{[100]}$ increases and saturates at a high value as the electrons are transferred to X which has a large mass and a low mobility along [100]. As expected, $R_{[010]}$ behaves opposite to $R_{[100]}$; [red trace in Fig. 2(d)].  Note that such a continuous valley tuning is not possible in other multivalley systems, such as Si \cite{Lai.2004} or single-layer graphene \cite{Feldman.2012, Feldman.2013}, although it is achievable via applying a perpendicular electric field in bilayer graphene \cite{Huang.2022}. 

We further tune the 2DES properties by mounting the sample on a rotatable stage and rotating it \textit{in situ} around [010], thus applying an in-plane magnetic field ($B_{||}$) along [100], as shown in Fig.  2(e).  For $\epsilon < 0$, the long axis of the occupied valley (X) is oriented parallel to $B_\|$, whereas for $\epsilon > 0$, when Y is occupied, this axis is oriented perpendicular to $B_\|$. We also performed self-consistent calculations of the charge distribution and electron Fermi sea at different  $B_{||}$ \cite{F1}.

%The effect of valley tuning can also be seen at high magnetic field data as shown in Fig. 2(d) demonstrating resistance anisotropy following $B=0$ data. Note that in all traces there is no FQHS at $\nu=1/2$; instead, the ground state is a compressible CF Fermi sea. The traces also reveal the presence of higher-order FQHSs, up to $\nu=7/15$ and $7/13$, attesting to the excellent sample quality.

Figure 3 demonstrates the evolution of the transport traces as we tilt the sample to introduce a $B_{||}$ component. The top panels (a) are for the case when the electrons occupy only the X valley. At $\theta=0^o$, there is no FQHS at $\nu=1/2$. As we tilt the sample, a FQHS develops at $\nu=1/2$ near $\theta \simeq 32^o$ as manifested by the deep minima in both $R_{[100]}$ and $R_{[010]}$ traces, and a plateau centered at $2h/e^2$ [see Fig. 1 for an enlarged version of the red trace in the center panel of Fig. 3 (a)]. At larger $\theta$, the $\nu=1/2$ FQHS weakens and is replaced by an insulating phase that raises both $R_{[100]}$ and $R_{[010]}$ at high $B_\perp$. This evolution is qualitatively similar to the one seen in 2DESs confined to wide GaAs QWs \cite{Hasdemir.PRB.2015}, and can be explained as follows. With increasing $\theta$, $B_{||}$ reduces the interlayer tunneling \cite{Hu.1992}. This can be seen from the calculated charge distributions shown in Fig. 3(a) insets. For an intermediate amount of tunneling, there is a FQHS at $\nu=1/2$, consistent with the findings of recent theories that predict a Pfaffian state in 2DESs confined to wide QWs with appropriate tunneling \cite{Zhu.PRB.2016, Faugno.PRL.2019, Zhao.PRB.2021}. At larger $\theta$, as the tunneling is further reduced, the $\nu=1/2$ FQHS is weakened, and is eventually engulfed by insulating phases that signal the formation of a bilayer Wigner crystal phase \cite{Manoharan.PRL.1996, Hasdemir.PRB.2015, Hatke.PRB.2017, fn.tilt}.

In Fig. 3(b) we show the evolution of the 2DES with $\theta$ when all the electrons are placed in the Y valley. We observe a qualitatively similar evolution, but with a notable exception: the $\nu=1/2$ FQHS is strong near $\theta \simeq 43^o$, much larger than $\theta \simeq 32^o$ observed for the X-valley case [Fig. 3(a)]. Note that in Fig. 3(a) the $1/2$ FQHS is already very weak at $\theta = 36.0^o$ and the insulating phase is setting in in full force. In contrast, for the Y-valley case in Fig. 3(b), at $\theta = 37.6^o$ the FQHS at $\nu=1/2$ has barely emerged. 

To highlight the difference between the X- and Y-valley evolutions, in Fig. 4 we summarize the relative strength of the $\nu=1/2$ FQHS as a function of $B_{||}$. As shown in Fig. 4(a) inset, we define the strength of the FQHS by the value of resistance at $\nu=1/2$ ($R_{1/2}$) normalized to the background resistance on the flanks of $\nu=1/2$ ($R_{bg}$). In Fig. 4(a) we show data for the ``hard axis,” i.e., data based on $R_{[100]}$ when the X valley is occupied [black traces in Fig. 3(a)], and $R_{[010]}$ when Y is occupied [the red traces in Fig. 3(b)]; these are shown in orange and green colors in Fig. 4(a), respectively. Data for the ``easy axis", i.e., based on red traces in Fig. 3(a) and black traces in Fig. 3(b), are shown in Fig. 4(b). In both Figs. 4(a) and (b), the $\nu=1/2$ FQHS is strongest at $B_{||} \simeq 8$ T when the electrons are in the X valley and at $B_{||}\simeq 12$ T when they are in Y.

It is clear in Figs. 3 and 4 that a larger $B_{||}$, or equivalently larger $\theta$, is required for the emergence of the $\nu=1/2$ FQHS in the Y-valley case compared to X. This can be explained by the fact that, in a quasi-2DES with finite (i.e., non-zero) electron layer thickness, the influence of an applied $B_{||}$ on the charge distribution and interlayer tunneling depends on the electron’s orbital motion and its effective mass in the direction perpendicular to $B_{||}$; see, e.g. Ref. \cite{Mueed.PRL.2015a}. Our calculated charge distributions, shown as insets in Fig. 3 plots, become more bilayer like, implying a lower interlayer tunneling, at a much smaller $\theta$ in (a) compared to (b). This is consistent with the appearance of the $\nu=1/2$ FQHS at smaller $\theta$ in (a), assuming that an appropriate (intermediate) amount of tunneling is required to observe the $1/2$ FQHS \cite{Zhu.PRB.2016, Faugno.PRL.2019, Zhao.PRB.2021}. 

%Treating $B_{||}$, applied along the $x$ direction, as a perturbative term in a single-particle, 1D Hamiltonian, we can write \cite{Gokmen.PRB.2008}:
%
%\begin{equation} \label{Eq1}
%H_{1D} = \frac{p_z^2}{2m_z}+ V(z)+ \frac{2 \hbar k_y qB_{||}z + (qB_{||}z)^2}{2m_y},
%\end{equation} 
%$m_x$, $m_y$, and $m_z$ are effective masses in $x$, $y$, and $z$ directions, $q$ is the electron charge, and $V(z)$ is the confinement potential in the $z$ direction. The perturbative part in the Hamiltonian has a term that is linear in $k_y$, and this causes a distortion of the Fermi contour and an enhancement of $m_y$, as seen in Fig. 4(e). 

A noteworthy observation in Figs. 3 and 4(a,b) is that the $1/2$ FQHS appears to be stronger when electrons are in the X valley; compare the orange data points to those in green in Figs. 4(a,b). We can further quantify this via measuring a ``pseudo energy gap" ($^{1/2}\Delta$) for the 1/2 FQHS as summarized in Figs. 4(c,d) \cite{Footnote.pseudo.gap, Ref.pseudo.gap}. We find that $^{1/2}\Delta \simeq 0.135$ K when the electrons are in the X valley [Fig. 4(c)], noticeably larger than $\simeq 0.085$ K for when they are in Y [Fig. 4(d)].  We can partly attribute this to how $B_{||}$ affects the Fermi sea and the effective mass of the quasi-2D electrons in our system (see Refs. \cite{Suppl.Mat., Gokmen.PRB.2008, Batke.PRB.1986, Kunze.PRB.1987}. From our self-consistent calculations \cite{F1}, we find mass anisotropy ratios $m_{[100]}/m_{[010]} \simeq 3.5$ for X electrons at $B_{||} = 7.7$ T and $m_{[010]}/m_{[100]} \simeq 8.5$ for Y electrons at $B_{||} = 11.7$ T where the 1/2 FQHS is strong. The much larger mass anisotropy for the Y-valley electrons might explain the smaller energy gap for the $1/2$ FQHS. Larger mass anisotropy is generally expected to weaken the FQHSs \cite{Wang.PRB.2012}, although the energy gaps could be quite robust and very large anisotropies would be needed to reduce the gaps significantly  \cite{Balram.PRB.2016, Jo.PRL.2017}.  Besides Fermi sea anisotropy, it is also likely that the different energy gaps measured in Figs. 4(c,d) result from the different electron charge distributions and tunneling for the different valley populations and different $B_{||}$ (Fig. 3 insets).

%From our self-consistent calculations \cite{F1}, we find $m_{[010]} \simeq 0.27m_0$ and $m_{[100]} \simeq 0.92m_0$ for X-valley electrons at $B_{||} = 7.7$ T and $m_{[010]} \simeq 1.4m_0$ and $m_{[100]} \simeq 0.17m_0$ for Y-valley electrons at $B_{||} = 11.7$ T where the 1/2 FQHS is strong. These imply mass anisotropy ratios $m_{[100]}/m_{[010]} \simeq 3.4$ for X electrons and $m_{[010]}/m_{[100]} \simeq 8.2$ for Y electrons. The much larger mass anisotropy for the Y-valley electrons might explain the smaller energy gap for the $\nu=1/2$ FQHS. Larger mass anisotropy is generally expected to weaken the FQHSs \cite{Wang.PRB.2012}, although the energy gaps could be quite robust and very large anisotropies would be needed to reduce the gaps significantly  \cite{Balram.PRB.2016, Jo.PRL.2017}.  Besides Fermi sea anisotropy, it is also likely that the different energy gaps measured in Figs. 4(c,d) result from the different electron charge distributions and tunneling for the different valley populations and different $B_{||}$ (Fig. 3 insets).

%Another possibility is the difference in CR effective mass, $\sqrt{m_x \times m_y} = 0.474m_0$ for Y electrons at 12 T and  $\sqrt{m_x \times m_y} = 0.496m_0$ for X electrons at 8 T. 

We also measured the evolution of the $1/2$ FQHS for the case where no in-plane strain is applied so that X and Y are equally occupied at $B=0$ [Fig. 2(d)]. The evolution, as detailed in the Supplemental Material \cite{Suppl.Mat.}, is similar to the case where only Y is occupied, i.e., the $1/2$ FQHS is strongest at $B_{||} \simeq 12$ T; see the blue data points in Figs. 4(a,b). This may appear surprising at first sight, but it can be readily explained based on the fact that, for a $B_{||}$ applied along the [100] direction, namely the long axis of X and short axis of Y, the Y-valley energy shifts to smaller values compared to the X valley \cite{Suppl.Mat., Gokmen.PRB.2008}. Again, the shift is related to how, in a quasi-2DES with finite layer thickness, the coupling of $B_{||}$ to the electrons' orbital motion and the resulting shift in energies and deformation of the charge distribution depend on the effective mass perpendicular to the direction of $B_{||}$ \cite{Gokmen.PRB.2008}.

Before closing, we emphasize that the calculation results shown in Fig. 3 insets should be interpreted cautiously. These are essentially Hartree calculations and ignore electron correlations. Moreover, they assume that $B_\perp =0$ \cite{F1}. As shown in Ref. \cite{Mueed.PRL.2015a}, at large $B_{||}$, the \textit{electron} Fermi sea in a quasi-2DES indeed splits and shows a bilayer behavior \cite{Suppl.Mat.}. In the presence of a large $B_\perp$, however, the measured Fermi sea for the CFs near $\nu=1/2$ remains connected at large $B_{||}$ and exhibits only moderate anisotropy \cite{Kamburov.PRB.2014}. This is true whether the ground state at $\nu=1/2$ is compressible \cite{Kamburov.PRB.2014}, or is an incompressible FQHS \cite{Mueed.PRL.2015b}.  In the case of an compressible, CF, ground state at $\nu=1/2$, the experimental finding of the connectivity of the CF Fermi sea has in fact been corroborated qualitatively by numerical, many-body calculations \cite{Ippoliti.PRB.2017}. This connectivity, as well as the presence of numerous one-component (odd-numerator) FQHSs such as $\nu=3/5, 5/9, 3/7,$ and $5/11$ on the nearby flanks of the $\nu=1/2$ FQHS provide strong evidence that the $1/2$ FQHS is likely also a one-component state, presumably a Pfaffian state as recent theories conclude \cite{Zhu.PRB.2016, Faugno.PRL.2019, Zhao.PRB.2021}. While we do not have an experimental measure of the shape or connectivity of the CF Fermi sea at large $B_{\perp}$ and $B_{||}$ in our sample, we do observe several odd-numerator FQHSs on the flanks of the $1/2$ FQHS (Fig. 1), similar to what is seen in 2D electron and hole systems confined to GaAs wide QWs \cite{Suen.PRL.1994, Shabani.PRB.2013, Liu.PRL.2014, Mueed.PRL.2015b, Mueed.PRL.2016}.

In summary, we observe transitions from a compressible CF phase to an incopressible FQHS to an insulating phase at $\nu=1/2$ as a function of increasing $B_{||}$ in a quasi-2DES confined to an AlAs QW with tunable valley occupancy and anisotropic Fermi-sea and effective-mass. We show that the transitions and the strength of $\nu=1/2$ FQHS depend strongly on the relative orientation of $B_{||}$ with respect to the axes of the occupied valley. The data can be explained qualitatively based on the coupling of $B_{||}$ to the orbital motion of the quasi-2D electrons, but a quantitative description awaits rigorous many-body calculations. Our results demonstrate a unique tuning of the even-denominator $\nu=1/2$ FQHS through controlling the valley occupancy and $B_{||}$.

\begin{acknowledgments} 
We acknowledge support through the U.S. Department of Energy Basic Energy Science (Grant No. DEFG02-00-ER45841) for measurements, and the National Science Foundation (Grants No.  DMR 2104771 and No. ECCS 1906253), the Eric and Wendy Schmidt Transformative Technology Fund, and the Gordon and Betty Moore Foundation’s EPiQS Initiative (Grant No. GBMF9615 to L. N. P.) for sample fabrication and characterization.  We also acknowledge QuantEmX travel grants from the Institute for Complex Adaptive Matter and the Gordon and Betty Moore Foundation through Grant No. GBMF5305 to M.S.H., M.K.M., and M.S. A portion of this work was performed at the National High Magnetic Field Laboratory, which is supported by the National Science Foundation Cooperative Agreement No. DMR-1644779 and the State of Florida. We thank S. Hannahs, T. Murphy, J. Park, H. Baek, and G. Jones at NHMFL for technical support. We also thank J. K. Jain for illuminating discussions.
\end{acknowledgments}


\begin{thebibliography}{99}

\bibitem {Jain.2007} J. K. Jain, \textit{Composite Fermions} (Cambridge University Press, New York, 2007).

\bibitem{Halperin.Jain.2020} B. I. Halperin, The Half-Full Landau Level,  in Fractional Quantum Hall Effects: New Developments, edited by B. I. Halperin and J. K. Jain (World Scientific Publishing Co., 2020); pp. 79-132.

\bibitem{Halperin.PRB.1993} B. I. Halperin, P. A. Lee, and N. Read, Theory of the half-filled Landau level, Phys. Rev. B \textbf{47}, 7312 (1993).

\bibitem {Willett.PRL.1987} R. L. Willett, J. P. Eisenstein, H. L. Stormer, D. C. Tsui, A. C. Gossard, and J. H. English, Observation of an even-denominator quantum number in the fractional quantum Hall effect. Phys. Rev. Lett. \textbf{59}, 1776 (1987).

\bibitem{Willett.RPP.2013} R. L. Willett, The quantum Hall effect at $5/2$ filling factor, Rep. Prog. Phys. \textbf{76}, 076501 (2013).

\bibitem{Falson.Nat.Phys.2015} J. Falson, D. Maryenko, B. Friess, D. Zhang, Y. Kozuka, A. Tsukazaki, J. H. Smet, and M. Kawasaki, Even-denominator fractional quantum Hall physics in ZnO, Nature Physics \textbf{11}, 347 (2015).

\bibitem{banerjee.2018} M. Banerjee, M. Heiblum, V. Umansky, D. E. Feldman, Y. Oreg, and A. Stern, Observation of half-integer thermal Hall conductance, Nature \textbf{559}, 205 (2018).

\bibitem {ShafayatAlAs.PRL.2018} Md. Shafayat Hossain, Meng K. Ma, Y. J. Chung, L. N. Pfeiffer, K. W. West, K. W. Baldwin, and M. Shayegan, Unconventional Anisotropic Even-Denominator Fractional Quantum Hall State in a System with Mass Anisotropy, Phys. Rev. Lett. \textbf{121}, 256601 (2018).

\bibitem{willett.2023} R. L. Willett, K. Shtengel, C. Nayak, L. N. Pfeiffer, Y. J. Chung, M. L. Peabody, K. W. Baldwin, and K. W. West, Interference Measurements of Non-Abelian $e/4$ \& Abelian $e/2$ Quasiparticle Braiding, Phys. Rev. X \textbf{13}, 011028 (2023).

\bibitem {Moore.Nucl.Phys.B.1991} G. Moore and N. Read, Nonabelions in the fractional quantum Hall effect. Nucl. Phys. B \textbf{360}, 362 (1991).

\bibitem {Nayak.Rev.Mod.Phys.2008} C. Nayak, S. H. Simon, A. Stern, M. Freedman, and S. Das Sarma, Non-abelian anyons and topological quantum computation, Rev. Mod. Phys. \textbf{80}, 1083 (2008).

\bibitem {Pan.SSC.2001} W. Pan, H. L. Stormer, D. C. Tsui, L. N. Pfeiffer, K. W. Baldwin, and K. W. West, Experimental evidence for a spin-polarized ground state in the $\nu=5/2$ fractional quantum Hall effect, Solid State Commun. \textbf{119} 641 (2001).

\bibitem {Tiemann.Science.2012} L. Tiemann, G. Gamez, N. Kumada, and K. Muraki, Unraveling the spin polarization of the $\nu=5/2$ fractional quantum Hall state, Science \textbf{335}, 6070 (2012).

\bibitem {Stern.PRL.2012} M. Stern, B. A. Piot, Y. Vardi, V. Umansky, P. Plochocka, D. K. Maude and I. Bar-Joseph, NMR probing of the spin polarization of the $\nu = 5/2$ quantum Hall state, Phys. Rev. Lett. \textbf{108} 066810 (2012).

\bibitem {Wurstbauer.PRL.2012} U. Wurstbauer, K. W. West, L. N. Pfeiffer, and A. Pinczuk, Phys. Rev. Lett. \textbf{110}, 026801 (2013).

\bibitem {Eisenstein.PRL.2017} J. P. Eisenstein, L. N. Pfeiffer, and K. W. West, Quantum Hall Spin Diode,
Phys. Rev. Lett. \textbf{118}, 186801 (2017).

\bibitem {Shafayat5/2.PRL.2018} Md. Shafayat Hossain, Meng K. Ma, M. A. Mueed, L. N. Pfeiffer, K. W. West, K. W. Baldwin, and M. Shayegan, Direct Observation of Composite Fermions and Their Fully-Spin-Polarized Fermi Sea near $\nu=5/2$, Phys. Rev. Lett. \textbf{120}, 256601 (2018).

\bibitem {Suen.PRL.1992} Y. W. Suen, L. W. Engel, M. B. Santos, M. Shayegan, and D. C. Tsui, Observation of a $\nu=1/2$ fractional quantum Hall state in a double-layer electron system, Phys. Rev. Lett. \textbf{68}, 1379 (1992).

\bibitem {Suen2.PRL.1992} Y. W. Suen, M. B. Santos, and M. Shayegan, Correlated states of an electron system in a wide quantum well, Phys. Rev. Lett. \textbf{69}, 3551 (1992). 

\bibitem {Suen.PRL.1994} Y. W. Suen, H. C. Manoharan, X. Ying, M. B. Santos, and M. Shayegan, Origin of the $\nu=1/2$ fractional quantum Hall state in wide single quantum wells, Phys. Rev. Lett. \textbf{72}, 3405 (1994).

\bibitem {Shabani.PRB.2013} J. Shabani, Yang Liu, M. Shayegan, L. N. Pfeiffer, K. W. West, and K. W. Baldwin, Phase diagrams for the stability of the $\nu=1/2$ fractional quantum Hall effect in electron systems confined to symmetric, wide GaAs quantum wells, Phys. Rev. B \textbf{88}, 245413 (2013).

\bibitem {Mueed.PRL.2015b} M. A. Mueed, D. Kamburov, S. Hasdemir, M. Shayegan, L. N. Pfeiffer, K. W. West, and K. W. Baldwin, Geometric Resonance of Composite Fermions Near the $\nu=1/2$ Fractional Quantum Hall State, Phys. Rev. Lett. \textbf{114}, 236406 (2015).

\bibitem {Mueed.PRL.2016}M. A. Mueed, D. Kamburov, L. N. Pfeiffer, K. W. West, K. W. Baldwin, and M. Shayegan, Geometric Resonance of Composite Fermions near Bilayer Quantum Hall States, Phys. Rev. Lett. \textbf{117}, 246801 (2016).

\bibitem {Liu.PRL.2014} Y. Liu, A. L. Graninger, S. Hasdemir, M. Shayegan, L. N. Pfeiffer, K. W. West, K. W. Baldwin, and R. Winkler, Fractional Quantum Hall Effect at $\nu=1/2$ in Hole Systems Confined to GaAs Quantum Wells, Phys. Rev. Lett. \textbf{112}, 046804 (2014).


\bibitem {Halperin.Helv.Phys.Acta.1983} B. I. Halperin, Helv. Phys. Acta \textbf{56}, 75 (1983).

\bibitem{Chakraborty.PRL.1987} T. Chakraborty and P. Pietil\"{a}inen, Fractional Quantum Hall Effect at Half-Filled Landau Level in a Multiple-Layer Electron System, Phys. Rev. Lett. \textbf{59}, 2784 (1987).

\bibitem{Yoshioka.PRB.1989} D. Yoshioka, A. H. MacDonald, and S. M. Girvin, Fractional quantum Hall effect in two-layered systems, Phys. Rev. B \textbf{39}, 1932 (1989).

\bibitem{Macdonald.SSc.1990} A. H. MacDonald, The fractional Hall effect in multi-component systems, Surf. Sci. \textbf{229}, 1 (1990).

\bibitem{He.PRB.1993} S. He, S. Das Sarma, and X. C. Xie, Quantized Hall effect and quantum phase transitions in coupled two-layer electron systems, Phys. Rev. B \textbf{47}, 4394 (1993)

\bibitem{Peterson.PRB.2010} M. R. Peterson and S. Das Sarma, Quantum Hall phase diagram of half-filled bilayers in the lowest and the second orbital Landau levels: Abelian versus non-Abelian incompressible fractional quantum Hall states, Phys. Rev. B \textbf{81}, 165304 (2010).

\bibitem{Thiebaut.PRB.2015} N. Thiebaut, N. Regnault, and M. O. Goerbig, Fractional quantum Hall states versus Wigner crystals in wide quantum wells in the half-filled lowest and second Landau levels, Phys. Rev. B \textbf{92}, 245401 (2015).

%\bibitem {Shabani.PRL.2009} J. Shabani, T. Gokmen, Y. T. Chiu, and M. Shayegan, Evidence for Developing Fractional Quantum Hall States at Even Denominator $1/2$ and $1/4$ Fillings in Asymmetric Wide Quantum Wells, Phys. Rev. Lett. \textbf{103}, 256802 (2009).

\bibitem {Zhu.PRB.2016} W. Zhu, Zhao Liu, F. D. M. Haldane, and D. N. Sheng, Fractional quantum Hall bilayers at half filling: Tunneling-driven non-Abelian phase, Phys. Rev. B 94, 245147 (2016).

\bibitem{Faugno.PRL.2019} W. N. Faugno, Ajit C. Balram, Maissam Barkeshli, and J. K. Jain, Prediction of a Non-Abelian Fractional Quantum Hall State with f -Wave Pairing of Composite Fermions in Wide Quantum Wells, Phys. Rev. Lett. \textbf{123}, 016802 (2019).

\bibitem{Zhao.PRB.2021} T. Zhao, W. N. Faugno, S. Pu, A. C. Balram, and J. K. Jain, Origin of the $\nu=1/2$ fractional quantum Hall effect in wide quantum wells, Phys. Rev. B \textbf{103}, 155306 (2021).

\bibitem {Greiter.PRB.1992} Martin Greiter, X. G. Wen, and Frank Wilczek, Paired Hall states in double-layer electron systems, Phys. Rev. B \textbf{46}, 9586 (1992).

\bibitem{Footnote.LL.crossing} FQHSs are also reported at $\nu=1/2$ in bilayer electron systems confined to GaAs double-quantum-well samples with negligible tunneling \cite{Eisenstein.PRL.1992} and in bilayer graphene \cite{Huang.2022}. In this case, the $\nu=1/2$ FQHS is likely a two-component, $\psi_{331}$ state. There are also FQHSs at $\nu=1/2$ at LL crossings in GaAs 2D hole systems \cite{Liu.PRB.2014} and graphene \cite{Zibrov.Nat.Phys.2018}.

\bibitem{Eisenstein.PRL.1992} J. P. Eisenstein, G. S. Boebinger, L. N. Pfeiffer, K. W. West, and Song He, New fractional quantum Hall state in double-layer two-dimensional electron systems, Phys. Rev. Lett. \textbf{68}, 1383 (1992).

\bibitem{Huang.2022} Ke Huang, Hailong Fu, Danielle Reifsnyder Hickey, Nasim Alem, Xi Lin, Kenji Watanabe, Takashi Taniguchi, and Jun Zhu, Valley Isospin Controlled Fractional Quantum Hall States in Bilayer Graphene, Phys. Rev. X \textbf{12}, 031019 (2022).

\bibitem{Liu.PRB.2014} Y. Liu, S. Hasdemir, D. Kamburov, A. L. Graninger, M. Shayegan, L. N. Pfeiffer, K. W. West, K. W. Baldwin, and R. Winkler, Even-denominator fractional quantum Hall effect at a Landau level crossing, Phys. Rev. B \textbf{89}, 165313 (2014).

\bibitem{Zibrov.Nat.Phys.2018} A. A. Zibrov, E. M. Spanton, H. Zhou, C. Kometter, T. Taniguchi, K. Watanabe, and A. F. Young, Even-denominator fractional quantum Hall states at an isospin transition in monolayer graphene, Nature Physics \textbf{14}, 930 (2018).

\bibitem{Lay} T. S. Lay, J. J. Heremans, Y. W. Suen, M. B. Santos, K. Hirakawa, and M. Shayegan, High‐quality two‐dimensional electron system confined in an AlAs quantum well, Appl. Phys. Lett. \textbf{62}, 3120 (1993).

\bibitem{Depoortere.APL.2002}E. P. De Poortere, Y. P. Shkolnikov, E. Tutuc, S. J. Papadakis, and M. Shayegan, Enhanced electron mobility and high order fractional quantum Hall states in AlAs quantum wells, Appl. Phys. Lett. \textbf{80}, 1583 (2002).

\bibitem{Shayegan.APL.2003}M. Shayegan, K. Karrai, Y. P. Shkolnikov, K. Vakili, E. P. De Poortere, and S. Manus, Low-temperature, in situ tunable, uniaxial stress measurements in semiconductors using a piezoelectric actuator, Appl. Phys. Lett. \textbf{83}, 5235 (2003).

\bibitem{Shayegan.AlAs.Review.2006} M. Shayegan, E. P. De Poortere, O. Gunawan, Y. P. Shkolnikov, E. Tutuc, and K. Vakili, Two-dimensional electrons occupying multiple valleys in AlAs, Phys. Stat. Sol. (b) \textbf{243}, 3629 (2006).

\bibitem {Chung.PRM.2018} Yoon Jang Chung, K. A. Villegas Rosales, H. Deng, K. W. Baldwin, K. W. West, M. Shayegan, and L. N. Pfeiffer, Multivalley two-dimensional electron system in an AlAs quantum well with mobility
exceeding $2\times10^{6}$ cm$^{2}$/Vs, Phys. Rev. Materials \textbf{2}, 071001 (R) (2018).

\bibitem {FN} Note that a finite $V_P$ is often required to attain $\varepsilon=0$ because of a sample- and cooldown-dependent residual strain \cite{Shayegan.AlAs.Review.2006}. 

\bibitem{Suppl.Mat.} See Supplemental Material for sample details, extended data, and supporting discussions.

\bibitem{Gokmen.Natphy.2010} T. Gokmen, Medini Padmanabhan, and M. Shayegan, Transference of transport anisotropy to composite fermions, Nature Physics \textbf{6}, 621 (2010).

\bibitem {Lai.2004} K. Lai, W. Pan, D. C. Tsui, S. Lyon, M. M\"{u}hlberger, and F. Sch\"{a}ffler, Two-Flux Composite Fermion Series of the Fractional Quantum Hall States in Strained Si, Phys. Rev. Lett. \textbf{93}, 156805 (2004).


\bibitem {Feldman.2012} B. E. Feldman, B. Krauss J.  H. Smet and A. Yacoby, Unconventional Sequence of Fractional Quantum Hall States in Suspended Graphene, Science \textbf{337}, 1196 (2012).

\bibitem {Feldman.2013} B. E. Feldman, A. J. Levin, B. Krauss, D. A. Abanin, B. I. Halperin, J. H. Smet, and A. Yacoby, Fractional Quantum Hall Phase Transitions and Four-Flux States in Graphene, Phys. Rev. Lett. \textbf{111}, 076802 (2013).


\bibitem {F1} Our self-consistent calculations of potential and charge distributions are carried out at $B=0$. However, since of interest here is the status of the 2DES at high magnetic fields near $\nu=1/2$ where the electrons presumably occupy the lowest, spin-polarized LL, we assume in all our self-consistent calculations that the 2DES is fully spin polarized and occupies only the lowest electric subband. We emphasize that a quantitative understanding of our experimental data requires rigorous, many-body calculations in the presence of both $B_{\perp}$ and $B_{||}$.

\bibitem {Hasdemir.PRB.2015} S. Hasdemir, Y. Liu, H. Deng, M. Shayegan, L. N. Pfeiffer, K. W. West, K. W. Baldwin, and R. Winkler, $\nu=1/2$ fractional quantum Hall effect in tilted magnetic fields, Phys. Rev. B \textbf{91}, 045113 (2015).

\bibitem {Hu.1992} J. Hu and A. H. MacDonald, Electronic structure of parallel two-dimensional electron systems in tilted magnetic fields, Phys. Rev. B \textbf{46}, 12554 (1992).

\bibitem{Manoharan.PRL.1996}H. C. Manoharan, Y. W. Suen, M. B. Santos, and M. Shayegan, Evidence for a Bilayer Quantum Wigner Solid, Phys. Rev. Lett. \textbf{77}, 1813 (1996).

\bibitem{Hatke.PRB.2017} A. T. Hatke, Yang Liu, L. W. Engel, L. N. Pfeiffer, K. W. West, K. W. Baldwin, and M. Shayegan, Microwave spectroscopic observation of a Wigner solid within the $\nu=  1/ 2$ fractional quantum Hall effect, Phys. Rev. B \textbf{95}, 045417 (2017).

\bibitem{fn.tilt}In wide QWs, via increasing the electron density and thus lowering the interlayer tunneling, the ground state at $\nu=1/2$ can be tuned from a compressible (CF) phase to an incompressible FQHS to an insulating phase which is likely a bilayer Wigner crystal state \cite{Suen2.PRL.1992, Suen.PRL.1994, Shabani.PRB.2013, Manoharan.PRL.1996, Hatke.PRB.2017}. This tuning can also be accomplished by keeping the density fixed and tilting the sample in the magnetic field, as is done here and also in Ref. \cite{Hasdemir.PRB.2015}.

%\bibitem {Liu.PRL.2012} Y. Liu, C. G. Pappas, M. Shayegan, L. N. Pfeiffer, K. W. West, and K. W. Baldwin, Observation of Reentrant Integer Quantum Hall States in the Lowest Landau Level, Phys. Rev. Lett. \textbf{109}, 036801 (2012).

\bibitem{Mueed.PRL.2015a} M. A. Mueed, D. Kamburov, M. Shayegan, L. N. Pfeiffer, K. W. West, K. W. Baldwin, and R. Winkler, Splitting of the Fermi Contour of Quasi-2D Electrons in Parallel Magnetic Fields, Phys. Rev. Lett. \textbf{114}, 236404 (2015).

\bibitem{Footnote.pseudo.gap} Because the resistance minimum at $\nu=1/2$ has a rising background on its flanks, we deduce a ``pseudo energy gap" \cite{Ref.pseudo.gap} by normalizing the resistance at $\nu=1/2$ to the average value of the flanking resistance peaks [see inset to Fig. 4(a)], and analyzing its activated behavior as shown in Figs. 4(c,d).

\bibitem {Ref.pseudo.gap}  J. R. Mallett, R. G. Clark, R. J. Nicholas, R. Willett, J. J. Harris, and C. T. Foxon, Experimental studies of the $\nu=1/5$ hierarchy in the fractional quantum Hall effect, Phys. Rev. B \textbf{38}, 2200(R) (1988).

\bibitem{Gokmen.PRB.2008}T. Gokmen, M. Padmanabhan, O. Gunawan, Y. P. Shkolnikov, K. Vakili, E. P. De Poortere, and M. Shayegan, Parallel magnetic-field tuning of valley splitting in AlAs two-dimensional electrons, Phys. Rev. B \textbf{78}, 233306 (2008).

\bibitem{Batke.PRB.1986} E. Batke and C. W. Tu, Effective mass of a space-charge layer on GaAs in a parallel magnetic field, Phys. Rev. B \textbf{34}, 3027 (1986).

\bibitem{Kunze.PRB.1987} U. Kunze, Effective-mass change of electrons in Si inversion layers under parallel magnetic fields, Phys. Rev. B \textbf{35}, 9168 (1987).

\bibitem{Wang.PRB.2012} H. Wang, R. Narayanan, X. Wan, and F. Zhang, Fractional quantum Hall states in two-dimensional electron systems with anisotropic interactions, Phys. Rev. B \textbf{86}, 035122 (2012).

\bibitem{Balram.PRB.2016} Ajit C. Balram and J. K. Jain, Exact results for model wave functions of anisotropic composite fermions in the fractional quantum Hall effect, Phys. Rev. B \textbf{93}, 075121 (2016).

\bibitem{Jo.PRL.2017} I. Jo, K. A. Villegas Rosales, M. A. Mueed, L. N. Pfeiffer, K. W. West, K. W. Baldwin, R. Winkler, M. Padmanabhan, and M. Shayegan, Transference of Fermi contour anisotropy to composite fermions, Phys. Rev. Lett. \textbf{119}, 016402 (2017).



%\bibitem{Suppl.Mat.} See Supplemental Material for sample details, extended data, and supporting discussions which includes Refs. \cite{Shayegan.APL.2003, Vakili.PRL.2004, Vakili.PRL.2005, Maryenko.PRL.2015, Gokmen.PRB.2010, Gunawan.PRL.2006, Koulakov.PRL.1996, Fogler.PRB.1996, Moessner.PRB.1996, Fradkin.PRB.1999, Fradkin.PRL.2000, Fradkin.ARCMP.2010, Lilly1.PRL.1999, Lilly.PRL.1999, Du.SSC.1999, Pan.PRL.1999, Shayegan.PhysicaE.2000, Manfra.PRL.2007,  Zhu.PRB.2017, Eisenstein.2014, Haldane.PRL.2011, Yang.PRB.2012, Abanin.PRB.2010, Kamburov.PRL.2013, Feldman.Science.2016, Xia.Natphys.2011, Koduvayur.PRL.2011, Liu.PRB.2013, MNK.PRB.2011, Maciejko.PRB.2013, Regnault.PRB.2017, Lee.Preprint.2018}.



%\bibitem{Vakili.PRL.2004} K. Vakili, Y. P. Shkolnikov, E. Tutuc, E. P. De Poortere, and M. Shayegan, Spin susceptibility of two-dimensional electrons in narrow AlAs quantum wells, Phys. Rev. Lett. \textbf{92}, 226401 (2004).
%
%\bibitem{Vakili.PRL.2005} K. Vakili, Y. P. Shkolnikov, E. Tutuc, N. C. Bishop, E. P. De Poortere, and M. Shayegan, Spin-dependent resistivity at transitions between integer quantum Hall states, \textit{Phys. Rev. Lett.} \textbf{94}, 176402 (2005).
%
%\bibitem{Maryenko.PRL.2015} D. Maryenko, J. Falson, M. S. Bahramy, I. A. Dmitriev, Y. Kozuka, A. Tsukazaki, and M. Kawasaki, Spin-selective electron quantum transport in nonmagnetic MgZnO/ZnO heterostructures, \textit{Phys. Rev. Lett.} \textbf{115} 197601 (2015).
%
%\bibitem{Gokmen.PRB.2010} T. Gokmen, M. Padmanabhan, and M. Shayegan, Contrast between spin and valley degrees of freedom, Phys. Rev. B \textbf{81}, 235305 (2010).
%
%\bibitem{Gunawan.PRL.2006} O. Gunawan, Y. P. Shkolnikov, K. Vakili, T. Gokmen, E. P. De Poortere, and M. Shayegan, Valley susceptibility of an interacting two-dimensional electron system, Phys. Rev. Lett. \textbf{97}, 186404 (2006).
%


%
%\bibitem {Abanin.PRB.2010}D. A. Abanin, S. A. Parameswaran, S. A. Kivelson, S. L. Sondhi, Nematic valley ordering in quantum Hall systems, Phys. Rev. B \textbf{82}, 035428 (2010).
%
%\bibitem {Kamburov.PRL.2013} D. Kamburov, Y. Liu, M. Shayegan, L. N. Pfeiffer, K. W. West, and K. W. Baldwin, Composite fermions with tunable Fermi contour anisotropy, Phys. Rev. Lett. \textbf{110}, 206801 (2013).
%
%
%\bibitem{Feldman.Science.2016} B. E. Feldman, M. T. Randeria, A. Gyenis, F. Wu, H. Ji, R. J. Cava, A. H. MacDonald, and A. Yazdani, Observation of a nematic quantum Hall liquid on the surface of bismuth, Science \textbf{354}, 316 (2016).
%
%
%\bibitem{Xia.Natphys.2011} J. Xia, J. P. Eisenstein, L. N. Pfeiffer, and K. W. West, Evidence for a fractionally quantized Hall state with anisotropic longitudinal transport, Nature Physics \textbf{7}, 845 (2011).
%
%\bibitem{Koduvayur.PRL.2011} S. P. Koduvayur, Y. Lyanda-Geller, S. Khlebnikov, G. Cs\'athy, M. J. Manfra, L. N. Pfeiffer, K. W. West, and L. P. Rokhinson, Effect of strain on stripe phases in the quantum Hall regime, Phys. Rev. Lett. \textbf{106}, 016804 (2011).
%
%\bibitem{Liu.PRB.2013} Y. Liu, S. Hasdemir, M. Shayegan, L. N. Pfeiffer, K. W. West, and K. W. Baldwin, Evidence for a $\nu=5/2$ fractional quantum Hall nematic state in parallel magnetic fields, Phys. Rev. B \textbf{88}, 035307 (2013).
%
%\bibitem {MNK.PRB.2011} M. Mulligan, C. Nayak, and S. Kachru, Effective field theory of fractional quantized Hall nematics, Phys. Rev. B \textbf{84}, 195124 (2011).
%
%\bibitem {Maciejko.PRB.2013} J. Maciejko, B. Hsu, S. A. Kivelson, Y. Park, and S. L. Sondhi, Field theory of the quantum Hall nematic transition, Phys. Rev. B \textbf{88}, 125137 (2013).
%
%\bibitem {Regnault.PRB.2017} R. Regnault, J. Maciejko, S. A. Kivelson, and S. L. Sondhi, Evidence of a fractional quantum Hall nematic phase in a microscopic model, Phys. Rev. B \textbf{96}, 035150 (2017).
%
%\bibitem{Lee.Preprint.2018} K. Lee, J. Shao, E. A. Kim, F. D. M. Haldane, and E. H. Rezayi, Pomeranchuk instability of composite Fermi liquids, Phys. Rev. Lett. \textbf{121}, 147601 (2018).

\bibitem {Kamburov.PRB.2014} D. Kamburov, M. A. Mueed, M. Shayegan, L. N. Pfeiffer, K. W. West, K. W. Baldwin, J. J. D. Lee, and R. Winkler, Fermi contour anisotropy of GaAs electron-flux composite fermions in parallel magnetic fields, Phys. Rev. B \textbf{89}, 085304 (2014).

%\bibitem {Footnote saying that the calculations are done at B_perp =0} The calculations are done at $B_\perp =0$.




%\bibitem {Manoharan.PRB.1994} H. C. Manoharan and M. Shayegan, Wigner crystal versus Hall insulator, \textit{Phys. Rev. B} \textbf{50}, 17662 (1994).
%
%
%\bibitem{He.PRB.1991} S. He, X. C. Xie, S. Das Sarma, and F. C. Zhang, Quantum Hall effect in double-quantum-well systems, Phys. Rev. B \textbf{43}, 9339 (1991).
%
%\bibitem {Li.Science.2017}J. I. A. Li, C. Tan, S. Chen, Y. Zeng, T. Taniguchi, K. Watanabe, J. Hone, and C. R. Dean, Even denominator fractional quantum Hall states in bilayer graphene, \textit{Science} \textbf{358} 648 (2017). 
%
%\bibitem {Zibrov.Nature.2017} A. A. Zibrov, C. Kometter, H. Zhou, E. M. Spanton, T. Taniguchi, K. Watanabe, M. P. Zaletel, and A. F. Young, Tunable interacting composite fermion phases in a half-filled bilayer-graphene Landau level, \textit{Nature} \textbf{549}, 360 (2017).
%
%
%
%\bibitem {Jain.PRL.1989} J. K. Jain, Composite-fermion approach for the fractional quantum Hall effect, Phys. Rev. Lett. \textbf{63}, 199 (1989).
%
%
%\bibitem {papic.PRB.2009} Z. Papi\'{c}, N. Regnault, and S. Das Sarma, Interaction-tuned compressible-to-incompressible phase transitions in quantum Hall systems, Phys. Rev. B \textbf{80}, 201303(R) (2009).
%
%
%\bibitem {Stern.PRL.1968} Frank Stern, Transverse Hall Effect in the Electric Quantum Limit, Phys. Rev. Lett. \textbf{21}, 1687 (1968).
%
%\bibitem{Tutuc.PRB.2003}E. Tutuc, S. Melinte, E. P. De Poortere, M. Shayegan, and R. Winkler, Role of finite layer thickness in spin polarization of GaAs two-dimensional electrons in strong parallel magnetic fields, Phys. Rev. B \textbf{67}, 241309(R) (2003).
%
%\bibitem{Halperin.1983} B. I. Halperin, Theory of the quantized Hall conductance, Helv. Phys. Acta 56, 75 (1983).
%
%
%\bibitem {Barkeshli.PRL.2018}M. Barkeshli, C. Nayak, Z. Papi\'c, A. F. Young, and M. Zaletel, Topological Exciton Fermi Surfaces in Two-Component Fractional Quantized Hall Insulators, Phys. Rev. Lett. 121, 026603 (2018).
%
%\bibitem {Zaletel.PRB.2018}M. P. Zaletel, S. Geraedts, Z. Papi\'{c}, and E. H. Rezayi, Evidence for a topological "exciton Fermi sea" in bilayer graphene, Phys. Rev. B 98, 045113 (2018).
%
%\bibitem {Wu.Nano.2017} Y.-H. Wu, T. Shi, and J. K. Jain, Non-Abelian Parton Fractional Quantum Hall Effect in Multilayer Graphene, Nano Lett., \textbf{17}, 4643 (2017).
%
%
%\bibitem {Kim.NatPhys.2019} Y. Kim, A. C. Balram, T. Taniguchi, K. Watanabe, J. K. Jain, J. H. Smet, Even denominator fractional quantum Hall states in higher Landau levels of graphene, Nature Physics \textbf{15}, 154 (2019).


%\bibitem{Jo.PRL.2017} Insun Jo, K. A. Villegas Rosales, M. A. Mueed, L. N. Pfeiffer, K. W. West, K. W. Baldwin, R. Winkler, Medini Padmanabhan, and M. Shayegan,  Transference of Fermi Contour Anisotropy to Composite Fermions,  Phys. Rev. Lett. \textbf{119}, 016402 (2017).



%\bibitem{Kamburov.PRB.2014} D. Kamburov, M. A. Mueed, M. Shayegan, L. N. Pfeiffer, K. W. West, K. W. Baldwin, J. J. D. Lee, and R. Winkler, Fermi contour anisotropy of GaAs electron-flux composite fermions in parallel magnetic fields, Phys. Rev. B \textbf{89}, 085304 (2014).

\bibitem{Ippoliti.PRB.2017} M. Ippoliti, Scott D. Geraedts, and R. N. Bhatt, Connection between Fermi contours of zero-field electrons and $\nu=1/2$ composite fermions in two-dimensional systems
Phys. Rev. B \textbf{96}, 045145 (2017).

\end{thebibliography}
\end{document}